\documentstyle[preprint,aps]{revtex}
\tightenlines
\begin{document}
\title{Reply to ``Comment on `Macroscopic Equation for the Roughness of
Growing Interfaces in Quenched Disorder' ''}
\author{L. A. Braunstein\cite{email},
R. C. Buceta and N. Giovambattista}
\address{Departmento de F\'{\i}sica, FCEyN, Universidad Nacional de Mar del
Plata, Funes 3350, Mar de Plata, Argentina}
\date{\today}
\maketitle


{\bf Braunstein {\sl et al.} replies:} In a comment of the recent
paper \cite{brauns}, Lopez {\sl et al.} \cite{espa} obtained
analytically the short time behavior of the directed percolation
depinning (DPD) model \cite{dpd}. Their results explains the
behavior of the temporal derivative of the interface width (DSIW)
for all $q$ until a time $t\simeq \mbox{e}^{-2}$. We argue that
the fail in reproducing the early time regime until the
correlations are generated ($t \simeq 1$ at the depinning
transition) is because the density of active sites of the
interface is not a constant $p$. This density depends on time as
we will show below. At time $t$ a site $i$, of a one dimensional
lattice of size $L$, is chosen at random with probability $1/L$.
Let us denote by $h_i(t)$ the height of the $i$-th generic site at
time $t$. The set of $\{h_i, i=1,\dots,L\}$ defines the interface
between ``wet'' and ``dry'' cells. We shall denote
$F_i=F_i(h_i+1)$ the activity of the $i$-th generic site above the
interface. If the cell $(i,h_i+1)$ is active $F_i=1$ (unblocked)
otherwise $F_i=0$ (blocked). The time evolution for the
probability of active sites $f(F_i=1,t)\equiv f(t)$ just above of
the interface in a time step $\delta t=1/L$ is
\begin{equation}
f(t+\delta t)=\frac{p}{L}\;f(t)+(1-\frac{1}{L})\; f(t)\label{q}\;.
\end{equation}
This equation takes into account the probability that a cell above
the interface remains active after a time step $\delta t$. The
first term take into account the probability of growth $f/L$ in
the $i$-th column times the probability $p$ that the new cell of
the interface be active. The second term is the probability that
no growth occurs in the $i$-th column when the cell is active.
Taking the limit $\delta t \to 0$ in Eq.~(\ref{q}) and solving the
equation, with initial condition $f(0)=p$, we obtain
\begin{equation}
f(t)=p\;\mbox{e}^{-q t}\label{f}\;.
\end{equation}
Notice that $f(t)$ is the interface activity density (IAD), {\sl
i.e.} $f(t)=\{\langle F_i \rangle\}$ where the brackets (braces)
denotes averages over the lattice (realizations). Notice that
$f(t)$ is close to $p$ until $t \simeq \mbox{e}^{-2}$ where the
result of L\'opez {\sl et. al.} holds. To obtain a more realistic
description for the DSIW until the correlation are generated ($t
\simeq 1$) it is necessary to take into account the temporal
dependence of the IAD. Let us consider a growth model in a system
of size $L$ with density $f(t)$ of active cells in the early time
regime. In this regimen the lateral correlations are negligible.
Assuming  independence between $F_i$ and $h_i$, the time evolution
for the probability of having a column with height $h$ at time $t$
is given by
\begin{equation}
P(h,t+ \delta t)\;=P(h-1,t) \frac {f(t)}{L} + P(h,t) \big(1-\frac
{f(t)}{L}\big)\label{prob}\;.
\end{equation}
Taking the limit $\delta t\to 0$ we obtain the master equation for
the probability. Using the generating function of moments, one can
calculate the DSIW:
\begin{equation}
\label{width} \frac{dw^2}{dt} = f(t)\;.
\end{equation}
For $t \gtrsim 1$ horizontal correlations are generated and
Eq.~(\ref{f}) breaks down. From numerical simulations we could
check that the hypothesis assumed to derive Eq~(\ref{prob}) holds
in the neighborhood of the criticality and in the pinned phase but
it breaks down for $q \ll q_c$. However, these values of $q$ are
not interesting from an experimental point of view
\cite{gene,brauns}. A comparison of Eq.(\ref{width}) with
numerical simulations of the DPD model \cite{dpd} is presented in
Fig.~\ref{gr} for the critical value. We can see that until $t
\lesssim \mbox{e}^{-2}$ the analytic result of Eq.~(\ref{width})
and the one obtained by L\'opez {\sl et al.} are coincident with
the numerical results of the DSIW. This is because in this regime
$f(t) \simeq p$. As time goes on, $f(t)$ decays and the hypothesis
of L\'opez {\sl et al.} does not hold as we can see in this
Figure. However, our analytical result predicts the DSIW until $t
\simeq 1$.

Finally, we argue that contrary to what it is claimed in the
comment of \cite{brauns}, Braunstein and Buceta's formula does
describe the macroscopic behavior of the interface even when the
solution is not exactly a matching between the early
[Eq.~(\ref{width})] and the asymptotic regimes.

\begin{figure}
\caption{Numerical results for the DSIW (circles) and the IAD
(triangles) for the DPD model in a system of size $L=8192$ for
$q_c = 0.539$. Continuous lines correspond to Eq.~(\ref{width})
and fit the data until $t \simeq 1$. Dashed line correspond to the
approximation of L\'opez {\sl et al.}. \label{gr}}
\end{figure}

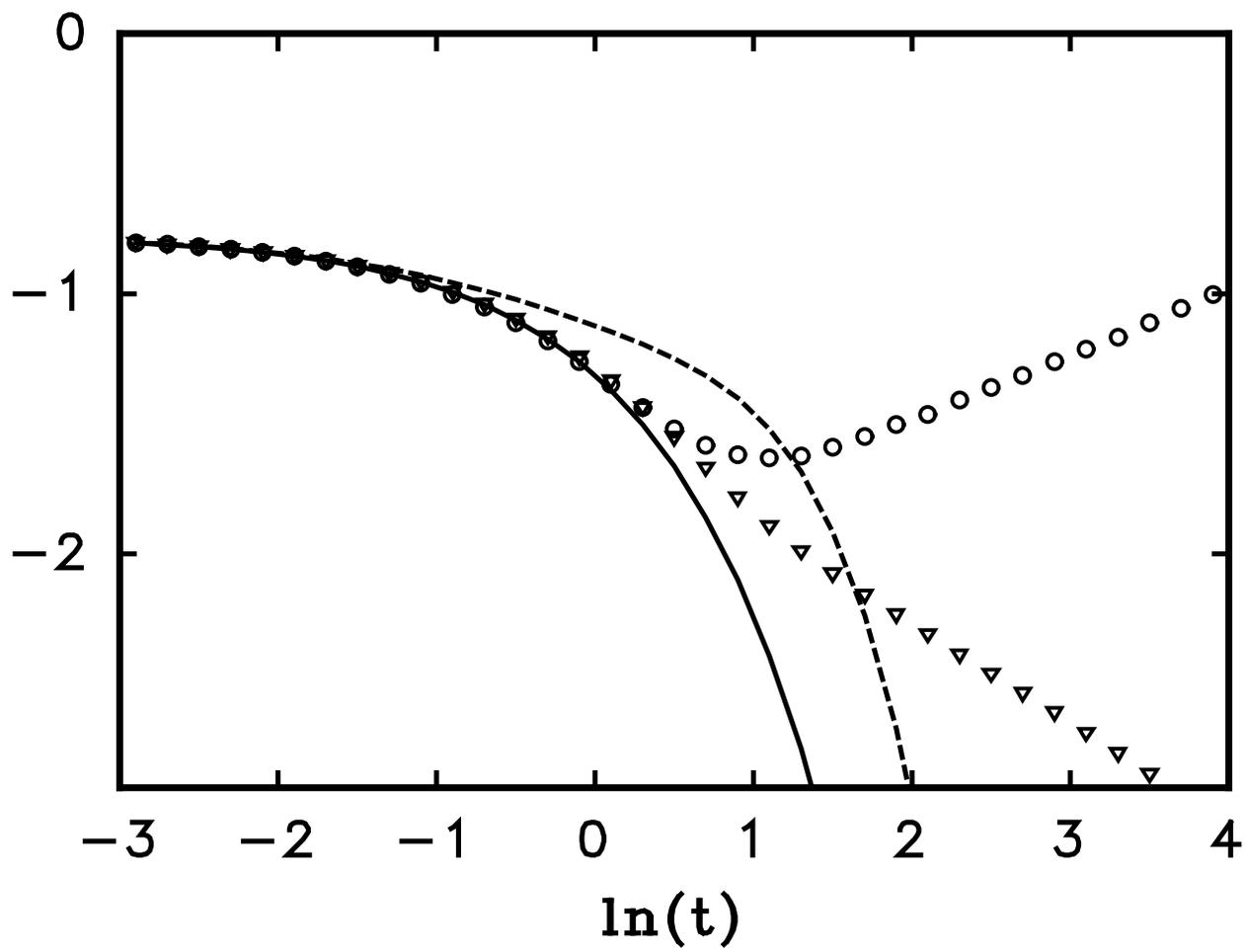
\end{document}